\documentclass[10pt,A4paper,conference]{IEEEtran}

\usepackage{graphics}
\usepackage[dvips]{graphicx}
\usepackage{epsfig}
\usepackage[dvips]{color}
\usepackage{amsfonts}
\usepackage{latexsym}
\usepackage{amsmath,amssymb}
\usepackage{psfrag}

\begin{document}

\title{Trellis Pruning for Peak-to-Average Power Ratio Reduction}
\author{\authorblockN{Mei Chen}
\authorblockA{Department of Electrical Engineering\\
University of Notre dame\\
Notre Dame, IN 46556 USA\\
Email: mchen1@nd.edu} \and
\authorblockN{Oliver M. Collins}
\authorblockA{Department of  Electrical Engineering\\
University of Notre dame\\
Notre Dame, IN 46556 USA\\
Email: ocollins@nd.edu}
 }

\maketitle

\begin{abstract}
This paper introduces a new trellis pruning method which uses
nonlinear convolutional coding for peak-to-average power ratio
(PAPR) reduction of filtered QPSK and 16-QAM modulations. The
 Nyquist filter is viewed as a convolutional
encoder that controls the analog waveforms of the filter output
directly. Pruning some edges of the encoder trellis can
effectively reduce the PAPR. The only tradeoff is a slightly lower
channel capacity and increased complexity. The paper presents
simulation results of the pruning action and the resulting PAPR,
and also discusses the decoding algorithm and the capacity of the
filtered and pruned QPSK and 16-QAM modulations on the AWGN
channel. Simulation results show that the pruning method reduces
the PAPR significantly without much damage to capacity.
\end{abstract}

\section{Introduction}
This paper presents a new coding-based method for peak-to-average
power ratio (PAPR) reduction of Nyquist filtered M-ary Phase-Shift
Keying (PSK) and M-ary Quadrature-Amplitude Modulation (QAM). Let
$x(t)$ denote the complex baseband signal after filtering and D/A
conversion in the transmitter. The PAPR is defined as
\begin{equation}\label{papr_def}
    \mathcal{P} = \frac{\max{|x(t)|^2}}{E[|x(t)|^2]},
\end{equation}
where $E[|x(t)|^2]$ is the expected value of $|x(t)|^2$
\cite{ochiaiCOMM2002}.

In literature, PAPR reduction is mostly addressed and considered
as an important issue for OFDM systems. However, it is a
substantial practical problem for Nyquist filtered modulations as
well. In a communication system link, due to bandwidth limitation,
a pulse shaping filter is required after baseband modulation to
suppress the sidelobes of the power spectra of PSK and QAM
signals. Root Raised Cosine (RRC) filters are commonly used as the
pulse shaping filter at the transmitter. And the receiver uses an
identical filter as matched filter so that the overall pulse
shaping spectrum will be the regular Raised Cosine (RC) spectrum
\cite{haykinWireless04}, which satisfies the Nyquist criterion for
distortionless baseband transmission in the absence of noise
\cite{haykinCOMMbook00}. The pulse shaping filter introduces
envelope variation to PSK and amplifies envelope fluctuation of
QAM. In the presence of envelope fluctuation, nonlinearity of the
transmit high-power amplifier (HPA) will cause intermodulation
distortion and the sidelobes which have been removed by filtering
will regrow. To prevent spectrum spreading, an HPA has to operate
in the linear zone, which leads to low power efficiency
\cite{liangHPA99}. Nonconstant envelope signals with pulse-shaping
filters (for example PSK and QAM) have better spectrum efficiency
than constant envelope signals. Yet people tend to use constant
envelope modulations such as GMSK \cite{murotaCOMM1981}, CPM
\cite{aulinCOMM81} and XPSK (also known as FQPSK)
\cite{feherCOMM83} to satisfy the power efficiency requirements.

A coding-based method can effectively reduce the PAPR of filtered
PSK and QAM signals and thus increases power efficiency. There are
many coding-based PAPR reduction methods for OFDM in the
literature, such as \cite{ahnVTC2000}, \cite{fanWCNC99}, and
\cite{wilkinsonVTC95}. However, these methods restrict the
application of coding to the digital domain. This paper presents a
new method employing a nonlinear convolutional coding strategy to
 control the analog waveform directly to achieve PAPR reduction.
The RRC filter is viewed as a convolutional encoder. Pruning
certain edges of the encoder trellis can eliminate output
waveforms that contribute to high PAPR. This trellis pruning
method can be generally applied to all PSK and QAM modulations.
QPSK and 16-QAM are considered as examples. Simulation results
show that the pruning method yields especially good results when
applied to 16-QAM, reducing the PAPR with negligible damage to
capacity. Section \ref{sec:Pruning Method} describes this new
method, and some PAPR simulation results are presented in Section
\ref{sec:Simulation Results}. Section \ref{sec:Capacity} discusses
the decoding algorithm and the capacity of the new restricted
channel. Section \ref{sec:Conclusion} concludes the paper.

\section{Trellis Pruning Method}\label{sec:Pruning Method}
\begin{figure}[h]
\centering
   \includegraphics[width=2in]{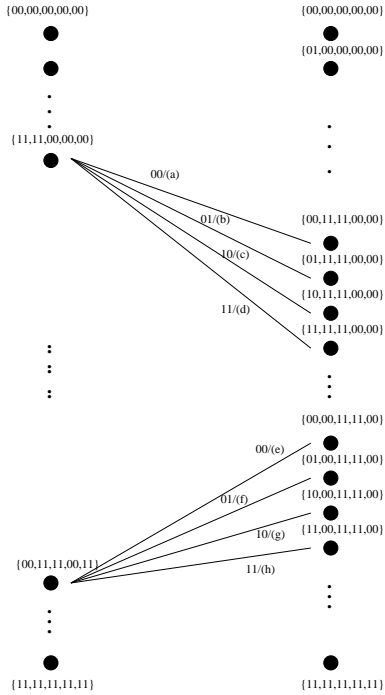}
   \caption{Trellis of an encoder with 1024 states}
    \label{fig:trellisnoprune}
\end{figure}

\begin{figure}[h]
\centering
   \includegraphics[width=3in]{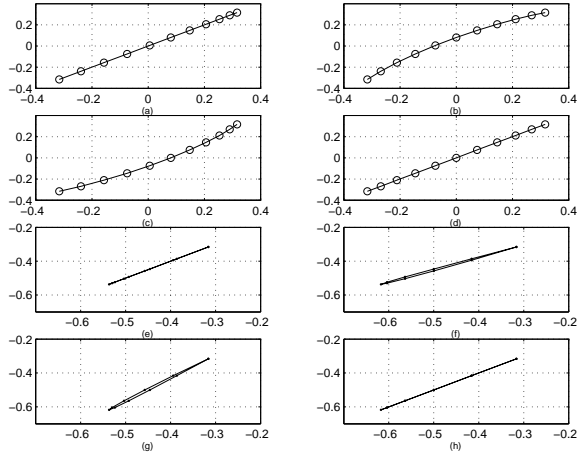}
   \caption{Output waveforms corresponding to the trellis shown in Fig. \ref{fig:trellisnoprune}}
    \label{fig:outputwave}
\end{figure}

\begin{figure}[h]
\centering
   \includegraphics[width=2in]{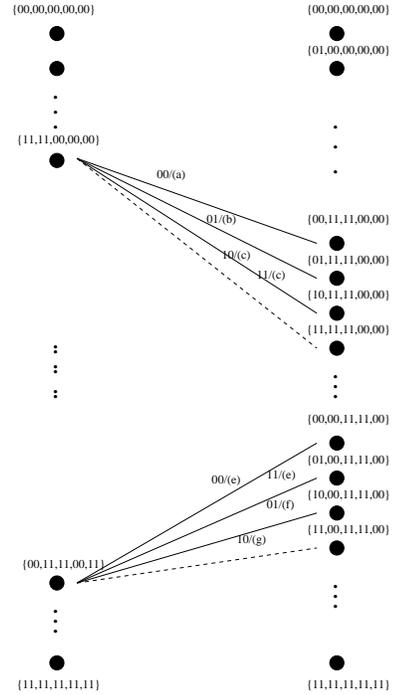}
   \caption{Pruned trellis (Pruned edges
are shown in dashed lines.)}
    \label{fig:trellisprune}
\end{figure}

Consider the root raised-cosine (RRC) pulse shaped QPSK
modulation. The RRC shaping pulse is
\begin{equation}\label{eqn:rrc}\begin{split}
    p(t) = \frac{\sqrt{2W}}{(1-(8\alpha Wt)^2)}\bigg(\frac{\sin(2\pi W(1-\alpha)t)}{2\pi
    Wt} \\
    + \frac{4\alpha}{\pi}\cos(2\pi W(1+\alpha)t)\bigg),
\end{split}\end{equation}
where $\alpha$ is the roll off factor and W is the ideal Nyquist
bandwidth \cite{haykinWireless04}. This paper considers a causal
pulse shaping filter with finite length given by
\begin{equation}\label{eqn:delayed_rrc}\begin{split}
    \hat{p}(t) = \begin{cases} p(t-DT), \text{ if } 0\leq t\leq 2DT \\
    0,\text{ otherwise}
\end{cases},
\end{split}\end{equation}
where $T$ is the symbol duration and the delay is $D$ symbols.
Truncating length of the filter is equal to twice the delay. The
filter output waveform is
\begin{equation}\label{eqn:output_waveform}
    s(t) = \sum_{i=0}^{\infty}g_i\hat p(t-iT),
\end{equation}
where $\{g_i\}$ is the input sequence with symbols drawn from QPSK
constellation.

We view the pulse shaping filter as a 4-ary convolutional encoder
with $2D-1$ shift registers. Therefore, total number of states of
the convolutional encoder is equal to $4^{2D-1}$. The output
associated with the state transition initiated by the $j$th input
QPSK message symbol $g_j$ is
\begin{eqnarray}
    s_j &=& s(t) \\
    &=& \sum_{i=0}^{\infty}g_i\hat p(t-iT) \\
    &=& \sum_{i=j-(2D-1)}^{j}g_i\hat p(t-iT)\text{, } \label{eqn:output_onesymbol}
\end{eqnarray}
for $jT\leq t \leq (j+1)T$. From \eqref{eqn:output_onesymbol},
$s_j$ is solely determined by the $j$th input and $2D-1$ message
symbols stored in the shift register. An example of the trellis of
such an encoder with 1024 states (D=3) is shown in Fig.
\ref{fig:trellisnoprune}. The corresponding output waveforms are
listed in Fig. \ref{fig:outputwave}.

In \eqref{eqn:output_onesymbol}, when $g_i\hat p(t-iT)$ adds
constructively, the trajectory of $s_j$ goes far away from the
origin; when $g_i\hat p(t-iT)$ adds destructively, the trajectory
of $s_j$ goes close to the origin. Therefore, some edges of the
trellis produce zero-crossing output waveforms (as is shown in
Fig. \ref{fig:outputwave} (a)-(d)), and some produce peaks (Fig.
\ref{fig:outputwave} (e)-(h)). Pruning such edges, i.e.,
eliminating the state transitions that generate large peak or
zero-crossing, reduces the PAPR significantly. Take the trellis in
Fig. \ref{fig:trellisnoprune} as example. Edge (d) and  (h) will
be pruned. The state transitions that follow the pruned edges will
take one of the three remaining edges from the same state. The new
trellis, which characterizes the pruning convolutional encoder, is
shown in Fig. \ref{fig:trellisprune}.

The trellis pruning method for QPSK modulation also applies to RRC
filtered 16-QAM modulation. The only difference is that the pulse
shaping filter is now viewed as a 16-ary convolutional encoder.

\section{PAPR Results}\label{sec:Simulation Results}

This section provides PAPR reduction results using the method
described in Section \ref{sec:Pruning Method}. There are many
pruning strategies to choose from. This paper uses the following
strategy: edges of the trellis that produce output waveforms going
farthest away from the origin (peak) and passing nearest to the
origin (zero-crossing) are simultaneously pruned. We use pruning
percentage $\eta$ to measure how much pruning is actually
performed. $\eta$ is defined  as
\begin{equation}\label{eqn:pruning_percentage}
    \eta = \frac{\text{number of pruned edges}}{\text{total number of edges in the
    trellis}}.
\end{equation}

\begin{figure}[h]
  \vspace{9pt}

  \centerline{\hbox{ \hspace{0.0in}
    \epsfxsize=1in
    \epsffile{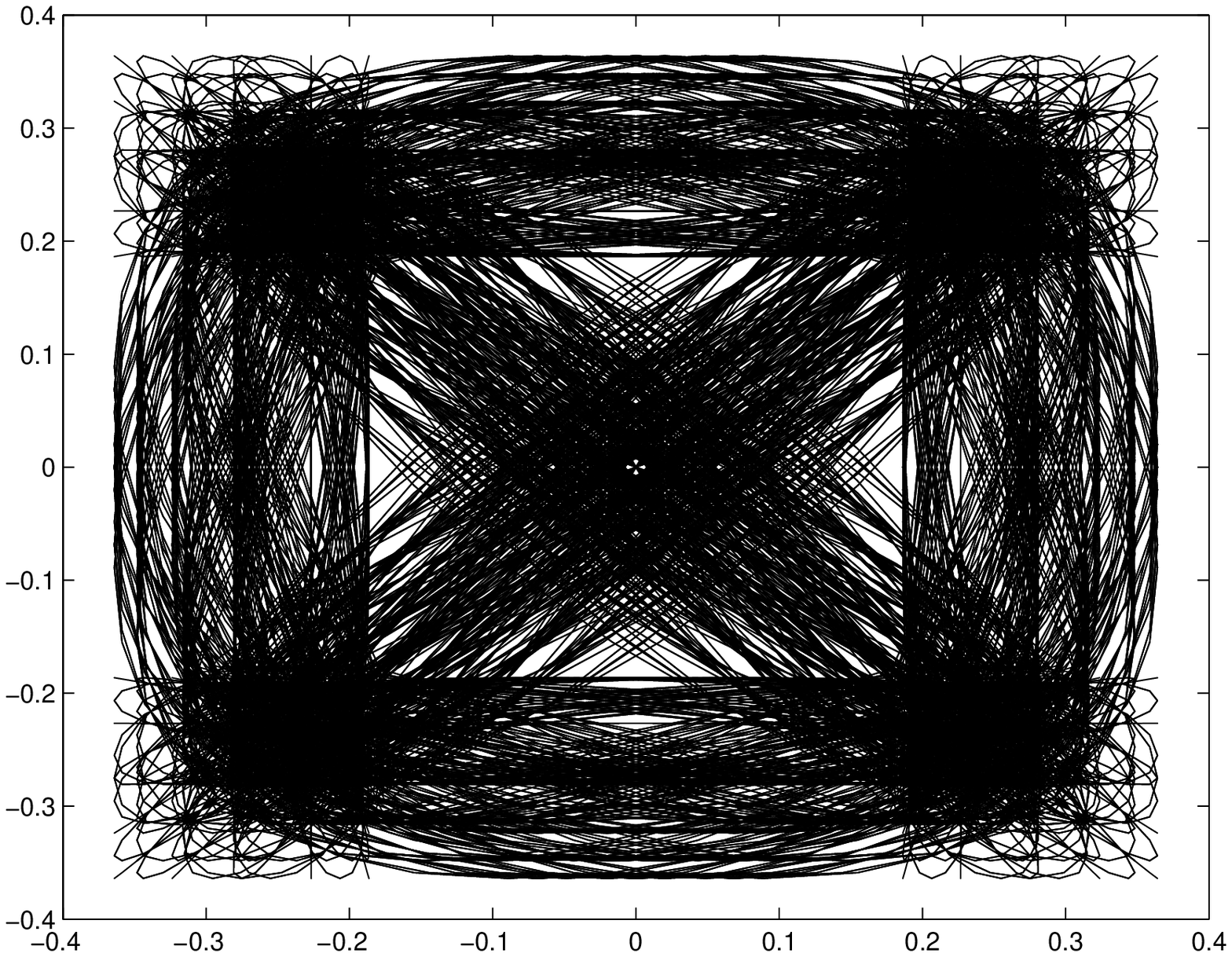}
    \hspace{0.25in}
    \epsfxsize=1in
    \epsffile{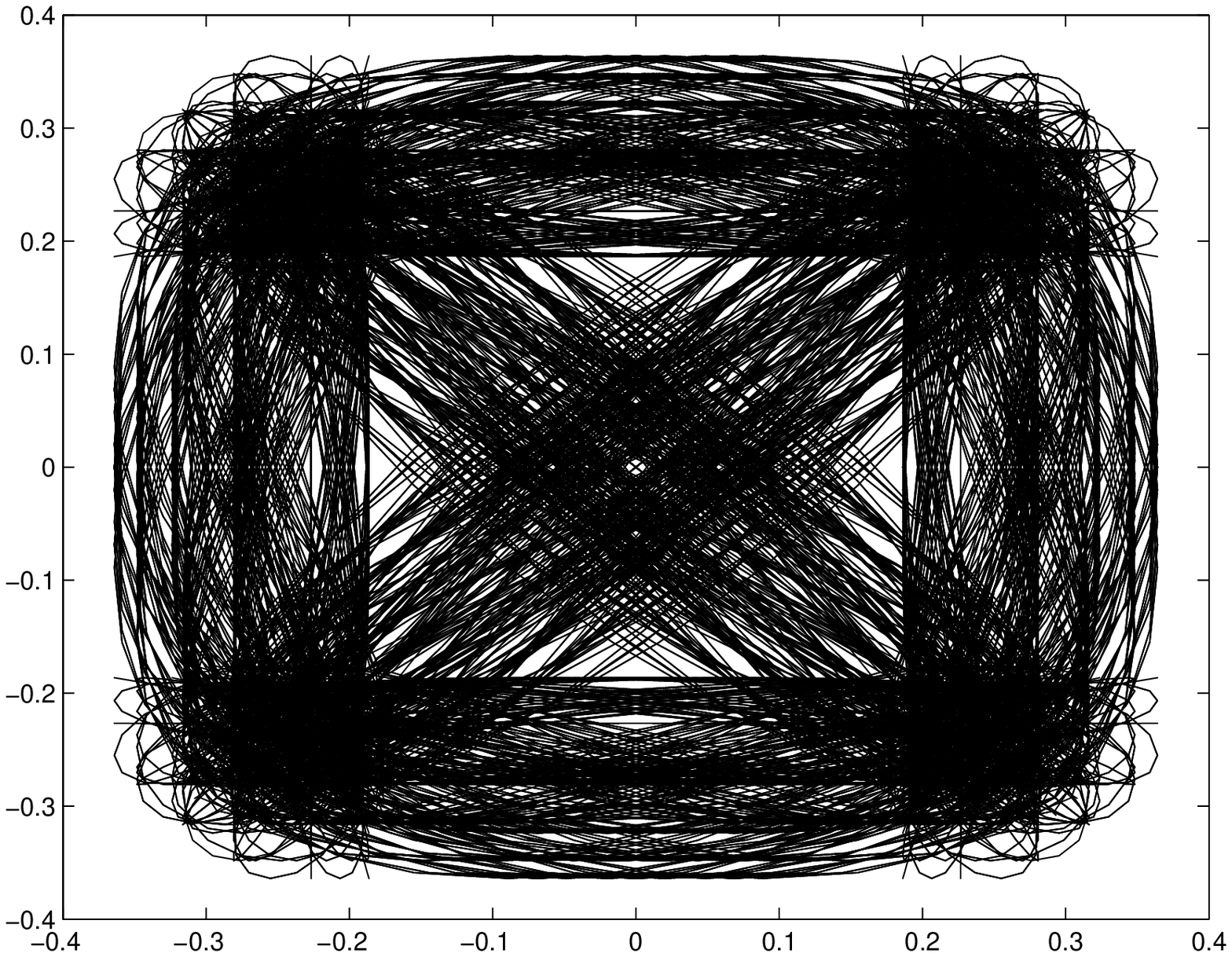}
    }
  }

  \vspace{9pt}
  \hbox{\hspace{1in} (a) \hspace{1.1in} (b)}
  \vspace{9pt}

  \centerline{\hbox{ \hspace{0.0in}
    \epsfxsize=1in
    \epsffile{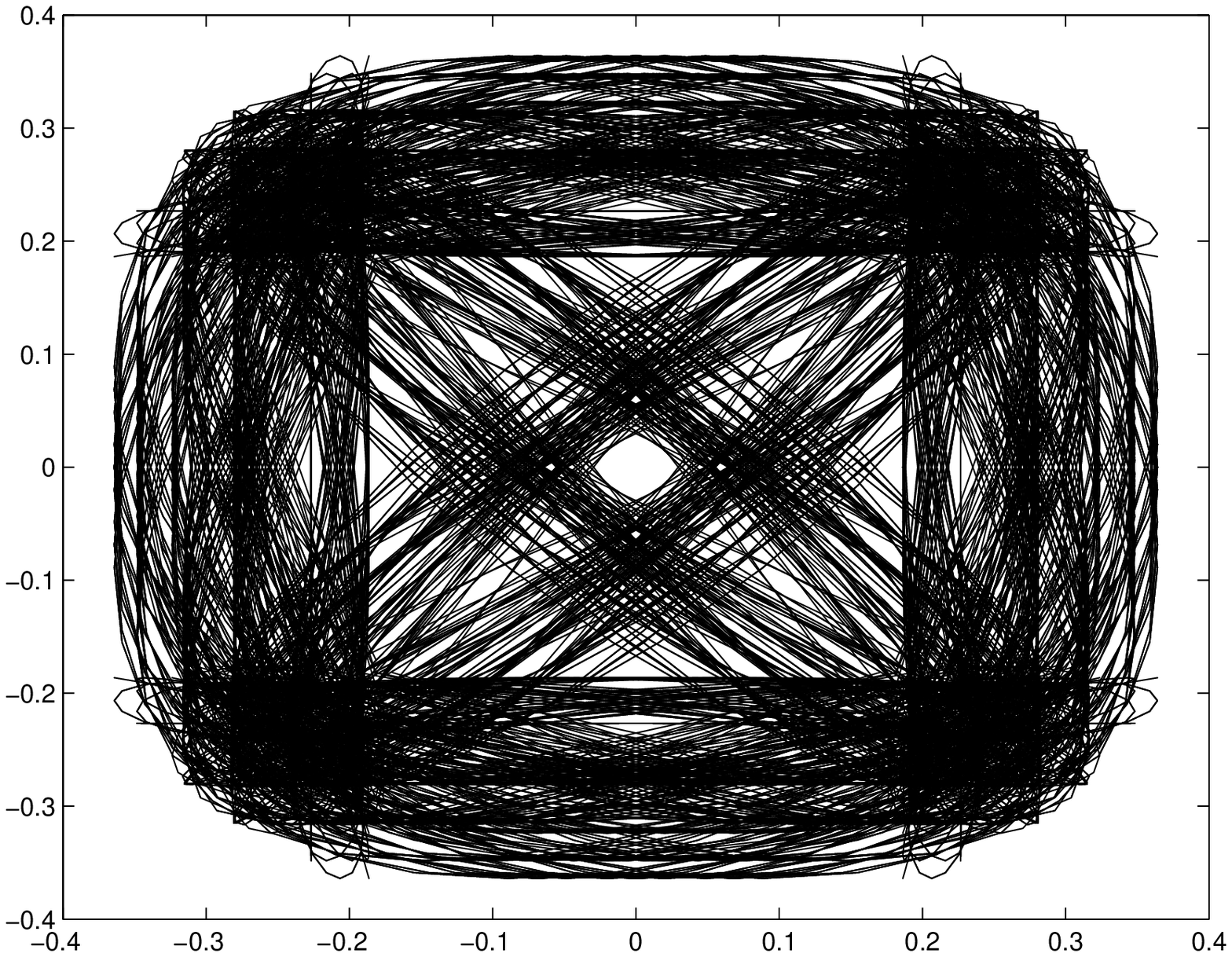}
    \hspace{0.25in}
    \epsfxsize=1in
    \epsffile{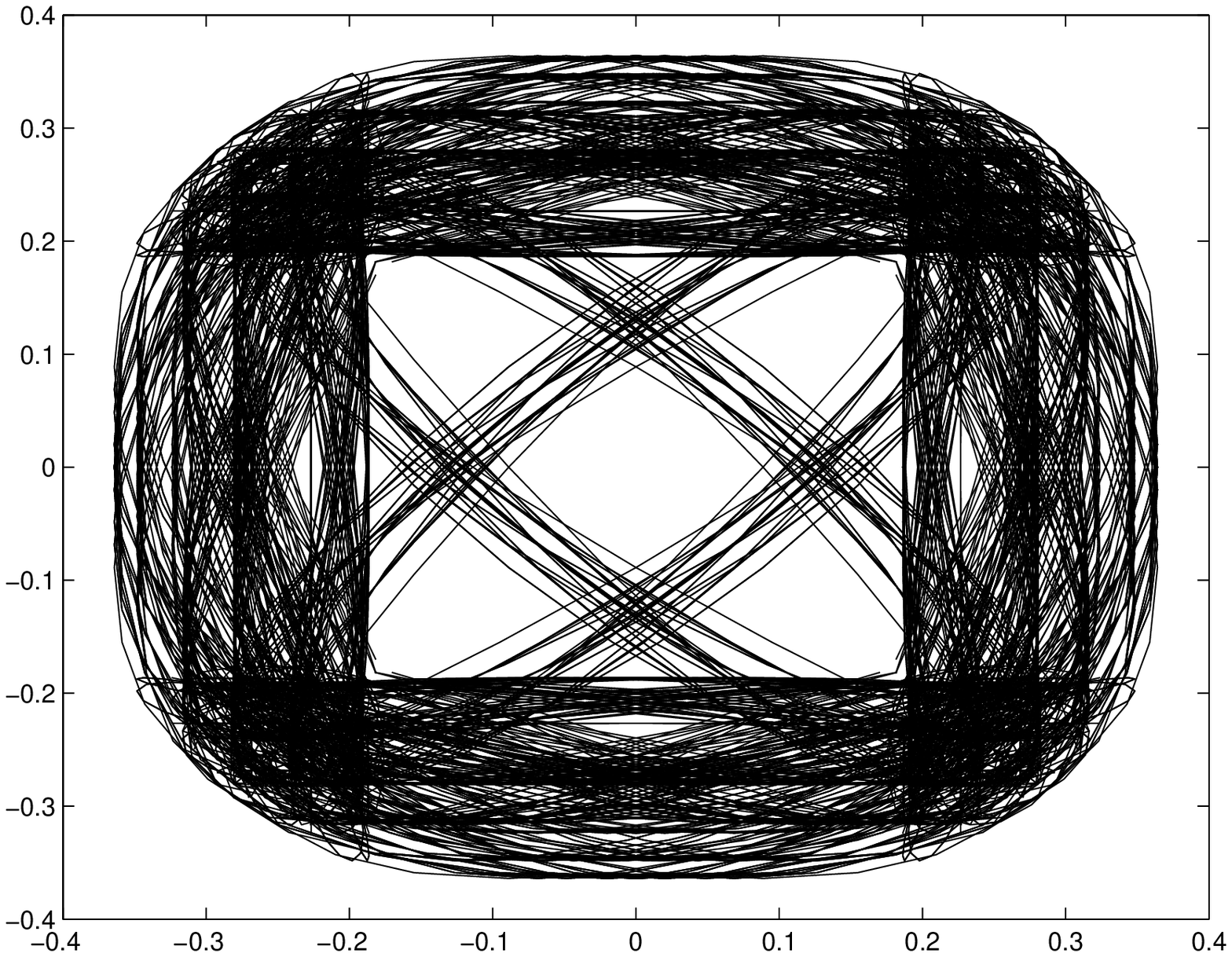}
    }
  }

  \vspace{9pt}
  \hbox{\hspace{1in} (c) \hspace{1.1in} (d)}
  \vspace{9pt}

  \caption{ Trajectory diagrams of RRC filtered QPSK (a)w/o pruning, (b)$10\%$ pruning, (c)$30\%$ pruning, and
(d)$50\%$ pruning.}
  \label{fig:constellation}

\end{figure}

All the filters used in the simulations have a delay $D$ of $3$
symbols. Fig. \ref{fig:constellation} shows the trajectory
diagrams of RRC filtered QPSK signals without pruning and with
$\eta = 10\%$, $30\%$, $50\%$ pruning respectively.  It is clear
that with the increase of the pruning percentage, zero-crossing
and peak waveforms are eliminated.

\begin{figure}[h]
\centering
   \includegraphics[width=3in]{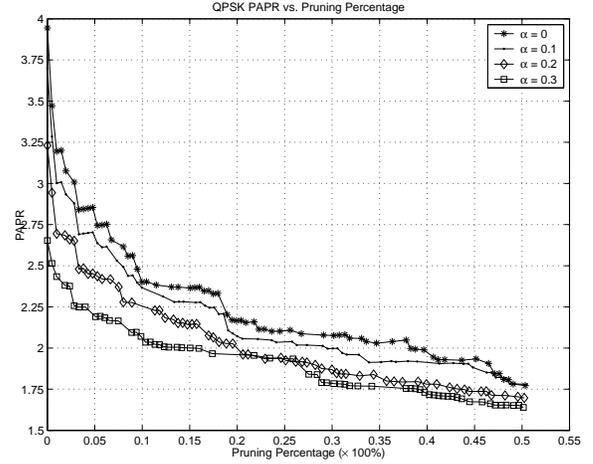}
   \caption{PAPR vs. pruning percentage for RRC filtered QPSK}
    \label{fig:papr}
\end{figure}

\begin{figure}[h]
\centering
   \includegraphics[width=3in]{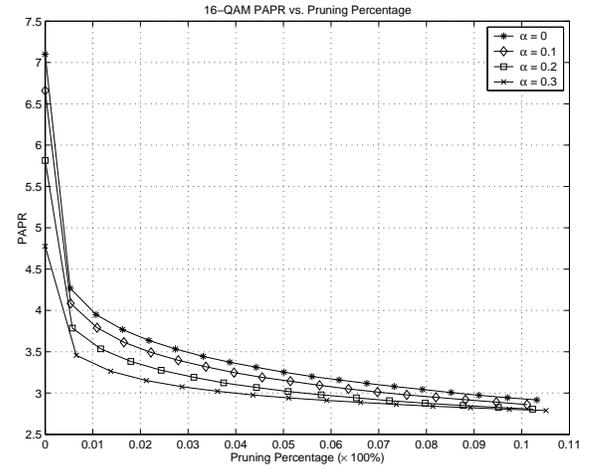}
   \caption{PAPR vs. pruning percentage for RRC filtered 16-QAM}
    \label{fig:qam_papr}
\end{figure}

The above pruning strategy is simulated for both QPSK and 16-QAM
for various pruning percentage, where the shaping filters are the
RRC filters of different roll-off factors. Fig. \ref{fig:papr} and
Fig. \ref{fig:qam_papr} shows that the pruning method can
effectively reduce the PAPR of both modulation schemes. The
reduction of the PAPR is more significant for 16-QAM:  a $0.5\%$
pruning can reduce the PAPR by more than $25\%$.

\section{Decoding Algorithm and Capacity}\label{sec:Capacity}
\subsection{Decoding Algorithm}\label{sec:bcjr}
Viterbi algorithm is typically used to decode convolutional codes.
However, the number of states of the encoder presented in this
paper grows exponentially with the length of the pulse shaping
filter and constellation size. With the increase of filter length
and the expansion of constellation from QPSK to 16-QAM, the
complexity of Viterbi algorithm makes it computationally
intractable. Therefore, we use a forward only decision feedback
aided BCJR algorithm with much lower complexity as the decoding
algorithm.

\begin{figure}[h]
\centering
    \begin{psfrags}
        \psfrag{x}{$\{X_k\}$}
        \psfrag{y}{$\{Y_k\}$}
        \includegraphics[width=3in]{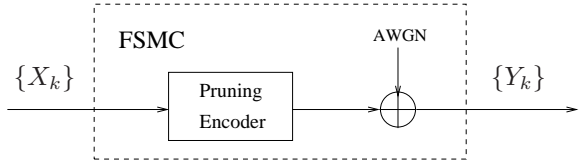}
    \caption{The pruning channel model}
    \label{fig:system}
    \end{psfrags}
\end{figure}
We model the pruning convolutional encoder as a finite-state
machine (FSM) \cite{costelloCodingbook}. The concatenation of the
FSM and the Additive White Gaussian Noise (AWGN) channel, as is
shown in Fig. \ref{fig:system}, is modelled as a Finite State
Markov Channel (FSMC). We call this particular FSMC \textit{the
pruning channel}. The channel input process $\{ X_k \}$ is assumed
to be i.i.d. symbols drawn from a finite set $\mathcal{S}$. The
state process $\{S_k\}$ forms an irreducible, aperiodic,
stationary finite state Markov chain over a finite state space
${\mathcal Q} = \{ 1, \cdots, {K} \}$.The statistics of the FSMC
satisfies the following properties:
\begin{itemize}
    \item $P(X_k|S_i) = P(X_k)$ for $i\leq k$;
    \item Given $X_k$ and $S_k$, $\{Y_k,S_{k+1}\}$ is statistically independent of
    $\{X_i\}_{i=1}^{k-1}$, $\{Y_i\}_{i=1}^{k-1}$, and $\{S_i\}_{i=1}^{k-1}$.
\end{itemize}

For FSMC, the standard BJCR \cite{bahlIT74} calculates  the
\textit{a posteriori} probability (APP) of $X_k$ given past and
future channel outputs $\{Y_i \}_{i = 1}^{N}$. Whereas, the
decision feedback aided BCJR (DFA-BCJR), proposed in
\cite{collinsISIT04} and \cite{liISIT05-1}, calculates the APP
value of $X_k$ given a finite sequence of the past channel inputs
$\{X_i\}_{i = 1}^{k-1}$ and a finite sequence of past and future
channel outputs $\{ Y_i \}_{i = 1}^{N}$. Denote $X_1^k \equiv
\{X_i \}_{i = 1}^{k}$ and $Y_1^k \equiv \{Y_i \}_{i = 1}^{k}$ for
notational convenience. From the Bayes rule,
\begin{equation}\label{eqn:APP-bayes}
P(X_k |Y_{1}^{N}, X^{k-1}_{1}) = \frac{P(Y_{1}^{N}|
X^{k}_{1})P(X_k)}{P(Y_{1}^{N}|X^{k-1}_{1})}.
\end{equation}
We can rewrite the conditional probabilities as
\begin{align}
& P(Y_{1}^{N}|X^{k}_{1}) =  \sum_{s'} \sum_{s}
\alpha_{k}(s')\gamma_{k+1}(s',s,X_k)
\beta_{k+1}(s), \label{eqn:bcjr_explicit} \\
& P(Y_{1}^{N}|X^{k-1}_{1}) = \sum_{s'} \sum_{s}
\alpha_{k}(s')\gamma_{k+1}(s',s) \beta_{k+1}(s),
\label{eqn:joint_P_2}
\end{align}
Where
\begin{align}
 &\alpha_{k}(s') = P(Y^{k-1}_{1}, S_{k}=s'|X_1^{k-1}),\label{eqn:alpha}\\
 &\gamma_{k+1}(s',s,X_k)  = P(Y_k,  S_{k+1}=s | S_{k}=s', X_k),\\
 &\gamma_{k+1}(s',s) = P(Y_k,  S_{k+1}=s | S_{k}= s'),\\
 &\beta_{k+1}(s)  = P(Y^{N}_{k+1}|S_{k+1}=s).
\end{align}
$\alpha_k$ denotes the forward path of the DFA-BCJR algorithm,
while $\beta_k$ denotes the backward path. DFA-BCJR algorithm
differs from the standard BCJR algorithm due to the presence of
known input symbols in the forward path. Note that when given
$X_1^{k-1}$, \eqref{eqn:alpha} can be rewritten as
\begin{equation}\label{eqn:APP-alpha-simplified}\begin{split}
    \alpha_k(s') = \begin{cases}1, \text{if } s' \text{ takes on value determined by } X_1^{k-1}\\
    0, \text{otherwise}
\end{cases}.
\end{split}\end{equation}
Substitute \eqref{eqn:APP-alpha-simplified} into
\eqref{eqn:bcjr_explicit} and \eqref{eqn:joint_P_2},
\begin{align}
& P(Y_{1}^{N}|X^{k}_{1}) =  \sum_{s} \gamma_{k+1}(s',s,X_k)
\beta_{k+1}(s), \label{eqn:cond-1} \\
& P(Y_{1}^{N}|X^{k-1}_{1}) = \sum_{s} \gamma_{k+1}(s',s)
\beta_{k+1}(s), \label{eqn:cond-2}
\end{align}
where the state $s'$ takes on value determined by $X_1^{k-1}$.

Equation \eqref{eqn:cond-1} and \eqref{eqn:cond-2} show that with
decision feedback, the complexity of the forward path calculation
is independent of the number of states $K$ of the FSMC. However,
the complexity of the backward path still grows exponentially with
$K$. To further reduce complexity, we only use the forward path of
the DFA-BCJR as the decoding algorithm. That is, when calculating
the APP value of $X_k$, no future channel outputs
$\{Y_k\}_{i=k+1}^{N}$ are available. Therefore, the APP value
becomes
\begin{equation}\label{eqn:APP-nofuture} P(X_k |Y_{1}^{k},
X^{k-1}_{1}) = \frac{ \sum_{s}
\gamma_{k+1}(s',s,X_k) }{ \sum_{s}
\gamma_{k+1}(s',s)}.
\end{equation}
From \eqref{eqn:APP-nofuture}, it is clear that the decoding
algorithm has a complexity growing linearly with the number of states.

To achieve good performance of the decoding algorithm, we need to
carefully choose the appropriate coding scheme before the
convolutional encoder, such that accurate estimation of the
previous input symbols can be obtained and then feedback to the
decoding algorithm as known input symbols.

\subsection{Capacity and Capacity Lower Bound}
This section discusses how to calculate the capacity and capacity
lower bound of the pruning channel. Simulation results show that
pruning does not decrease capacity much. Here capacity refers to
the constrained channel capacity of a given input distribution. We
only consider $M$-ary i.i.d. inputs.

The capacity of the pruning channel, which is modelled as an FSMC
shown in Fig. \ref{fig:system}, is
\begin{equation}\label{eqn:cap}
    C = \lim_{N\rightarrow\infty}\frac{1}{N}I({X_1^N};{Y_1^N}).
\end{equation}
Expanding $\frac{1}{N}I(X^N_1;Y^N_1)$, we have
\begin{align}
  \frac{1}{N}&I(X^N_1;Y^N_1)= \frac{1}{N}[H(X^N_1)-H(X^N_1|Y^N_1)] \nonumber\\
   &= \log_2M -\frac{1}{N}\sum_{k=1}^N H(X_k|Y^N_1,X_1^{k-1})\label{eqn:entropy1}\\
   &= \log_2M -\frac{1}{N}\sum_{k=1}^N
   E[-\log_2P(X_k|Y_1^N,X_1^{k-1})]. \label{eqn:exact_capacity}
\end{align}

As  $N$ goes to infinity, \eqref{eqn:exact_capacity} is the
capacity of the pruning channel. $P(X_k|Y_1^N,X_1^{k-1})$ in
\eqref{eqn:exact_capacity} can be computed by the DFA-BCJR
algorithm. Therefore, we can calculate the channel capacity by
Monte Carlo simulation of equation \eqref{eqn:APP-bayes} to
\eqref{eqn:cond-2} and \eqref{eqn:exact_capacity}.

The computation becomes too complex when the filter length or $M$
is large. To simplify calculation, we derive a capacity lower
bound as follows. Since conditioning reduces entropy, we have
\begin{equation}\label{eqn:inequality}
    H(X_k|Y^N_1,X_1^{k-1}) \leq H(X_k|Y_1^k,X_1^{k-1}).
\end{equation}
Substituting \eqref{eqn:inequality} into \eqref{eqn:entropy1}
yields
\begin{align}
  \frac{1}{N}&I(X^N_1;Y^N_1) \\
   &\geq \log_2M -\frac{1}{N}\sum_{k=1}^N H(X_k|Y_1^k,X_1^{k-1})\label{eqn:entropy2}\\
   &= \log_2M -\frac{1}{N}\sum_{k=1}^N
   E[-\log_2P(X_k|Y_1^k,X_1^{k-1})]. \label{eqn:low_bound}
\end{align}
Equation \eqref{eqn:low_bound} is the capacity lower bound of the
pruning channel. It can be simulated using equation
\eqref{eqn:APP-nofuture}. This lower bound is also the achievable
rate of the forward only DFA-BCJR decoder discussed in Section
\ref{sec:bcjr}.

\begin{figure}[h]
\centering
   \includegraphics[width=2.9in]{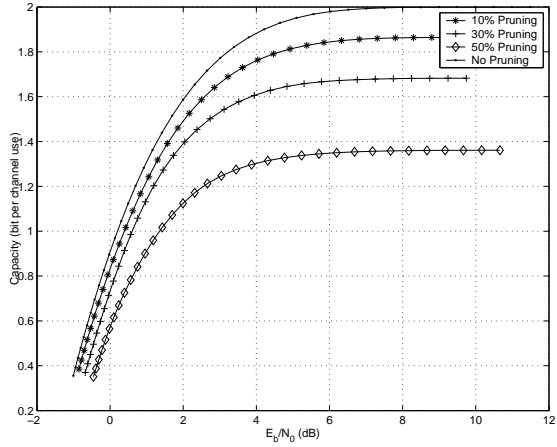}
   \caption{Capacity of the AWGN channel and the $10\%$, $30\%$, $50\%$ pruning channel with QPSK inputs}
    \label{fig:qpsk_capacity}
\end{figure}

\begin{figure}[h]
\centering
   \includegraphics[width=2.9in]{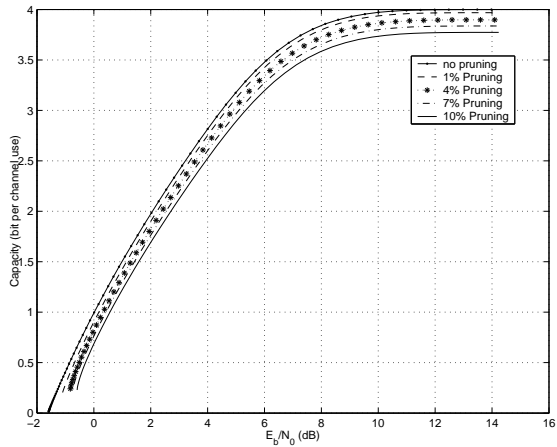}
   \caption{Capacity of the AWGN channel and capacity lower bound of the $1\%$, $4\%$, $7\%$ and $10\%$ pruning channel with 16-QAM inputs}
    \label{fig:qam_capacity}
\end{figure}
\begin{figure}[h]
\centering
   \includegraphics[width=2.9in]{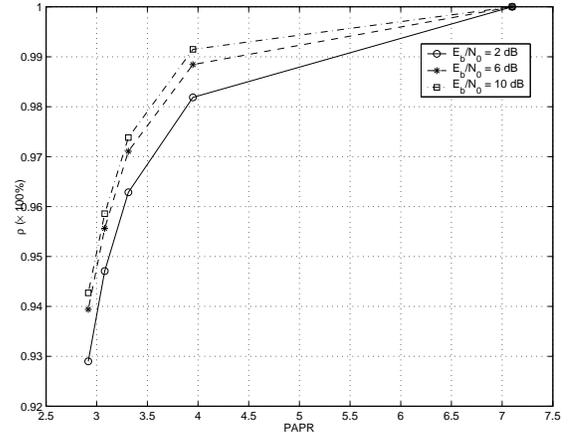}
   \caption{Percentage of capacity remaining after pruning vs. PAPR for RRC filtered 16-QAM signals}
    \label{fig:qam_loss}
\end{figure}

Capacity of the pruning channel with QPSK inputs and capacity
lower bound of the pruning channel with 16-QAM inputs are shown in
Fig. \ref{fig:qpsk_capacity} and Fig. \ref{fig:qam_capacity}
respectively. These results illustrate that, capacity loss,
compared with the normal AWGN channel without pruning, is fairly
small when pruning is less than $10\%$. For 16-QAM, $10\%$ pruning
reduces the PAPR by more than a half. Define
\begin{equation}\label{eqn:remaining_capacity}
    \rho = \frac{\text{Capacity of the pruning channel}}{\text{Capacity of the normal AWGN channel}}
\end{equation}
to represent the amount of capacity that is preserved after
pruning. Fig. \ref{fig:qam_loss}, which combines Fig.
\ref{fig:qam_papr} and Fig. \ref{fig:qam_capacity}, further
confirms that with a small loss of capacity, the pruning method
can achieve significant PAPR reduction for filtered 16-QAM
modulation.

\section{Conclusion}\label{sec:Conclusion}

This paper introduces a new trellis pruning method for reducing
the PAPR of filtered QPSK and 16-QAM signals. The pulse shaping
filter is viewed as a nonlinear convolutional encoder. Eliminating
certain state transitions of the encoder can reduce the PAPR of
transmitted signals. Simulation results confirm that this method
can significantly reduce the PAPR of both QPSK and 16-QAM
modulations.

Forward only DFA-BCJR algorithm is used as the decoding algorithm.
DFA-BCJR algorithm also helps to calculate the capacity of the
pruning channel. Capacity loss is small when pruning is less than
$10\%$, which means for 16-QAM, reducing the PAPR by more than a
half only costs a negligible capacity loss.

\bibliographystyle{IEEEbib}
{
\bibliography{qpskpapr}}

\end{document}